# Spinning Fast Iterative Data Flows

Stephan Ewen [1] Kostas Tzoumas [2] Moritz Kaufmann [3]
Volker Markl [4]

Technische Universität Berlin, Germany
[1,2,4] firstname.lastname@tu-berlin.de [3] moritz.kaufmann@campus.tu-berlin.de

## ABSTRACT

Parallel dataflow systems are a central part of most analytic pipelines for big data. The iterative nature of many analysis and machine learning algorithms, however, is still a challenge for current systems. While certain types of bulk iterative algorithms are supported by novel dataflow frameworks, these systems cannot exploit computational dependencies present in many algorithms, such as graph algorithms. As a result, these algorithms are inefficiently executed and have led to specialized systems based on other paradigms, such as message passing or shared memory.

We propose a method to integrate *incremental iterations*, a form of workset iterations, with parallel dataflows. After showing how to integrate bulk iterations into a dataflow system and its optimizer, we present an extension to the programming model for incremental iterations. The extension alleviates for the lack of *mutable state* in dataflows and allows for exploiting the *sparse computational dependencies* inherent in many iterative algorithms. The evaluation of a prototypical implementation shows that those aspects lead to up to two orders of magnitude speedup in algorithm runtime, when exploited. In our experiments, the improved dataflow system is highly competitive with specialized systems while maintaining a transparent and unified dataflow abstraction.

## 1. INTRODUCTION

Parallel dataflow systems are an increasingly popular solution for analyzing large data volumes. They offer a simple programming abstraction based on directed acyclic graphs, and relieve the programmer from dealing with the complicated tasks of scheduling computation, transfering intermediate results, and dealing with failures. Most importantly, they allow dataflow programs to be distributed across large numbers of machines, which is imperative when dealing with today's data volumes. Besides parallel databases [17, 19], MapReduce [16] is the best known representative, popular for its applicability beyond relational data. Several other systems, like Dryad [23], Hyracks [11], and Stratosphere [7], follow that trend and push the paradigm further, eliminating many shortcomings of MapReduce.

While dataflow systems were originally built for tasks like indexing, filtering, transforming, or aggregating data, their simple interface and powerful abstraction have made them popular for other kinds of applications, like machine learning [5] or graph analysis [26]. Many of these algorithms are of *iterative* or *recursive* nature, repeating some computation until a condition is fulfilled. Naturally, these tasks pose a challenge to dataflow systems, as the flow of data is no longer acyclic.

During the last years, a number of solutions to specify and execute iterative algorithms as dataflows have appeared. MapReduce extensions like *Twister* [18] or *HaLoop* [13], and frameworks like *Spark* [36] are able to efficiently execute a certain class of iterative algorithms. However, many machine learning and graph algorithms still perform poorly, due to those systems' inability to exploit the (sparse) computational dependencies present in these tasks [28]. We refer to the recomputed state as the partial solution of the iteration, and henceforth distinguish between two different kinds of iterations:

- *Bulk Iterations*: Each iteration computes a completely new partial solution from the previous iteration's result, optionally using additional data sets that remain constant in the course of the iteration. Prominent examples are machine learning algorithms like Batch Gradient Descend [35] and Distributed Stochastic Gradient Descent [37], many clustering algorithms (such as K-Means), and the well known PageRank algorithm[1].
- *Incremental Iterations*: Each iteration's result differs only partially from the result of the previous iteration. Sparse computational dependencies exist between the elements in the partial solution: an update on one element has a direct impact only on a small number of other elements, such that different parts of the solution may converge at different speeds. An example is the Connected Components algorithm, where a change in a vertex's component membership directly influences only the membership of its neighbors. Further algorithms in this category are many graph algorithms where nodes propagate changes to neighbors, such as shortest paths, belief propagation, and finding densely connected sub-components. For certain algorithms, the updates can be applied asynchronously, eliminating the synchronization barrier between iterations.

Existing iterative dataflow systems support bulk iterations, because those iterations resemble the systems' batch processing mode: the algorithms fully consume the previous iteration's result and compute a completely new result. In contrast, incrementally iterative algorithms evolve the result by changing or adding some data points, instead of fully recomputing it in a batch. This implies updating a mutable state that is carried to the next iteration. Since existing dataflow systems execute incremental iterations as if they were bulk iterative, they are drastically outperformed by specialized systems [28, 29].

[1] We refer to the original batch version of the PageRank algorithm. An incremental version of the algorithm exists [25].

Existing dataflow systems are therefore practically inefficient for many iterative algorithms. The systems are, however, still required for other typical analysis and transformation tasks. Hence, many data processing pipelines span multiple different systems, using workflow frameworks to orchestrate the various steps. Training a model over a large data corpus frequently requires a dataflow (like MapReduce) for preprocessing the data (e. g., for joining different sources and normalization), a specialized system for the training algorithm, followed by another dataflow for postprocessing (such as applying the model to assess its quality) [35].

We argue that the integration of iterations with dataflows, rather than the creation of specialized systems, is important for several reasons: first, an integrated approach enables many analytical pipelines to be expressed in a unified fashion, eliminating the need for an orchestration framework. Second, dataflows have been long known to lend themselves well to optimization, not only in database systems, but also when using more flexible programming models [7,22]. Third, dataflows seem to be a well adopted abstraction for distributed algorithms, as shown by their increased popularity in the database and machine learning community [5, 35].

The contributions of this paper are the following:

- We discuss how to integrate bulk iterations in a parallel dataflow system, as well as the consequences for the optimizer and execution engine (Section 4).
- We discuss an *incremental iteration* abstraction using *worksets*. The abstraction integrates well with the dataflow programming paradigm, can exploit the inherent computational dependencies between data elements, allowing for very efficient execution of many graph and machine learning algorithms (Section 5).
- We implement bulk and incremental iterations in the Stratosphere system, and integrate iterative processing with Stratosphere's optimizer and execution engine.
- We present a case study, comparing the performance of graph algorithms in a state-of-the-art batch processing system, a dedicated graph analysis system, and our own Stratosphere dataflow system that supports both bulk and incremental iterations. Our experimental results indicate that incremental iterations are competitive with the specialized system, outperforming both the batch sytem and Stratosphere's own bulk iterations by up to two orders of magnitude. At the same time, Stratosphere outperforms the specialized system for bulk iterations (Section 6).

The remaining sections are structured as follows. Section 2 reviews general concepts of iterative computations. Section 3 recapitulates the basic features of dataflow systems that we assume in the course of this paper. Section 7 discusses related work, and Section 8 concludes and offers an outlook.

# 2. ITERATIVE COMPUTATIONS

This section recapitulates the fundamentals of iterations and different representations that lend themselves to optimized execution.

## 2.1 Fixpoint Iterations

An iteration is, in its most general form, a computation that repeatedly evaluates a function $f$ on a partial solution $s$ until a certain termination criterion $t$ is met:

**while** $\neg t(s, f(s))$ **do**
  $s = f(s)$

A specific class of iterations are fixpoint computations, which apply the step function until the partial solution no longer changes:

**while** $s \neq f(s)$ **do**
  $s = f(s)$

For continuous domains, the termination criterion typically checks whether a certain error threshold has been achieved, rather than exact equality: $t(s, f(s)) \equiv (|s - f(s)| \leq \epsilon)$.

Fixpoint iterations compute the Kleene chain of partial solutions $(s, f(s), f^2(s), \ldots, f^i(s))$, and terminate when $f^k(s) = f^{k+1}(s)$ for some $k > 0$. The value $k$ is the number of iterations needed to reach the fixpoint $f^k(s)$. Denote $s_i = f^i(s)$. Fixpoint iterations are guaranteed to converge if it is possible to define a *complete partial order* (CPO) $\preceq$ for the data type of $s$, with a bottom element $\perp$. Furthermore, the step function $f$ must guarantee the production of a successor to $s$ when applied: $\forall s : f(s) \succeq s$. The existence of a supremum and the guaranteed progress towards the supremum result in eventual termination.

*Example: Connected Components.* Assume an undirected graph $G = (V, E)$. We wish to partition the vertex set $V$ into maximal subsets $V_i \subseteq V$ such that all vertices in the same subset are mutually reachable. We compute a solution as a mapping $s : V \to \mathbb{N}$, which assigns to each vertex a unique number (called component ID) representing the connected component the vertex belongs to: $\forall v \in V_i, w \in V_j : s(v) = s(w) \Leftrightarrow i = j$. Algorithm FIXPOINT-CC in Table 1 shows pseudocode of the pure fixpoint implementation of the Connected Components algorithm. The algorithm takes as input the set of vertices $V$ and a neighborhood mapping $N : V \to V^*$, which assigns to each vertex the set of its immediate neighbors: $\forall x, v \in V : x \in N(v) \Leftrightarrow (v, x) \in E \vee (x, v) \in E$. The mapping $s$ is the partial solution and is iteratively improved. Initially, $s(v)$ is a unique natural number for each vertex $v$ (we can simply number the vertices from 1 to $|V|$ in any order). Line 2 of the algorithm corresponds to the termination condition $s \prec f(s)$, and lines 3-5 correspond to the partial solution update $s \leftarrow f(s)$: For each vertex, its component ID is set to the minimal component ID of itself and all its neighbors. Like all algorithms in the second column of Table 1, FIXPOINT-CC returns $s$ as the result of the iteration.

The CPO over $s$ is defined by comparing the component IDs assigned to vertices: $s \succeq s' \Leftrightarrow \forall v \in V : s(v) \leq s'(v)$. A simple supremum is the mapping that assigns zero to all vertices.

## 2.2 Incremental Iterations & Microsteps

For many fixpoint algorithms, the partial solution $s$ is a set of data points and the algorithms do not fully recompute $s_{i+1}$ from $s_i$, but rather update $s_i$ by adding or updating some of its data points. Frequently, the change to a data point in one iteration affects only few other data points in the next iteration. For example, in most algorithms that operate on graphs, changing a vertex immediately affects its neighbors only. This pattern is often referred to as *sparse computational dependencies* [28, 29].

To support such algorithms efficiently, we can express an iteration using two distinct functions $u$ and $\delta$, instead of a single step function $f$. Algorithm INCR of Table 1 provides pseudocode for this iteration scheme. The $\delta$ function computes the *working set* $w = \delta(s, f(s))$, which is conceptually the set of (candidate) updates that, when applied to $s$, produce the next partial solution. The function $u$ combines $w$ with $s$ to build the next partial solution: $f(s) = u(s, w)$. Because the evaluation of $f(s)$ is what we seek to avoid, we vary this pattern to evaluate $\delta$ on $s_i$ and $w_i$ to compute the next working set $w_{i+1}$.

The function $u$ is typically efficient when $w$ contains only candidate updates relevant to the current iteration. Consequently, this form of *incremental* iterations is of particular interest, if a $\delta$ function exists that is both effective, in that it adds only relevant candidate updates to $w$, and efficient, in that it does not require the evaluation of $f(s)$. Incremental iterations are similar to workset algorithms used in optimizing compilers when performing dataflow analysis [24]. To the best of our knowledge, no formal characterization of the class of functions whose fixpoint computation is amenable to such an optimization exists. It is known that minimal fixpoints of all distributive functions $f : T \to T : f(A \vee B) = f(A) \vee f(B)$, where $T$ is a semilattice with a meet operator $\vee$, can be executed in an incremental fashion [14].

| Iteration Template | Connected Components |
|---|---|
| 1: **function** FIXPOINT($f$, $s$)<br>2: **while** $s \prec f(s)$ **do**<br>3: $s = f(s)$ | 1: **function** FIXPOINT-CC($V$,$N$)<br>2: **while** $(\exists v, x \in V \| x \in N(v) \wedge s(x) < s(v))$ **do**<br>3: **for** $(v \in V)$ **do**<br>4: $m = \min\{s(x) \| x \in N(v)\}$<br>5: $s(v) = \min\{m, s(v)\}$ |
| 1: **function** INCR($\delta$,$u$,$s$,$w$)<br>2: **while** $w \neq \emptyset$ **do**<br>3: $w' = \delta(s, w)$<br>4: $s = u(s, w)$<br>5: $w = w'$ | 1: **function** INCR-CC($V$,$N$)<br>2: **while** $w \neq \emptyset$ **do**<br>3: $w' = \emptyset$<br>4: **for** $(x, c) \in w$ **do**<br>5: **if** $c < s(x)$ **then**<br>6: **for** $z \in N(x)$ **do**<br>7: $w' = w' \cup \{(z, c)\}$<br>8: **for** $(x, c) \in w$ **do**<br>9: **if** $c < s(x)$ **then**<br>10: $s(x) = c$<br>11: $w = w'$ |
| 1: **function** MICRO($\delta$,$u$,$s$,$w$)<br>2: **while** $w \neq \emptyset$ **do**<br>3: $d = \mathbf{arb}(w)$<br>4: $s = u(s, d)$<br>5: $w = w \cup \delta(s, d)$ | 1: **function** MICRO-CC($V$,$N$)<br>2: **while** $w \neq \emptyset$ **do**<br>3: $(d, c) = \mathbf{arb}(w)$<br>4: **if** $(c < s(d))$ **then**<br>5: $s(d) = c$<br>6: **for** $(z \in N(d))$ **do**<br>7: $w = w \cup \{(z, c)\}$ |

**Table 1: Classes of iterations and the corresponding implementations of the Connected Components algorithm. The arb function selects and removes an arbitrary element from a set.**

*Example: Connected Components.* Algorithm INCR-CC in Table 1 shows pseudocode for the incremental implementation of the Connected Components algorithm. The working set $w$ contains in each step the new candidate component IDs for a set of vertices. Initially, $w$ consists of all pairs $(v, c)$ where $c$ is the component ID of a neighbor of $v$. For each vertex that gets a new component ID, $\delta$ adds this ID as a candidate for all of the vertex's neighbors (lines 4-7). The function $u$ updates the partial solution in lines 8-10 of the algorithm. For each element of the working set, it replaces the vertex's component ID by a candidate component ID, if the latter is lower. This representation implicitly exploits the computational dependencies present in the algorithm: a vertex's component can only change if one of its neighbors' component changed in the previous iteration. In the algorithm, a new candidate component is only in the working set for exactly those vertices.

In practice, one can frequently obtain an effective and efficient $\delta$ function by decomposing the iterations into a series of *microsteps*, and eliminating the *supersteps*. A microstep removes a single element $d$ from $w$ and uses it to update $s$ and $w$, effectively interleaving the updates of the partial solution and the working set. Microstep iterations lead to a modified chain of solutions ($s \preceq p_{0,1} \preceq \ldots \preceq p_{0,n} \preceq f(s) \preceq p_{1,1} \preceq \ldots \preceq p_{1,n} \preceq f^2(s), \ldots$), where $p_{i,j}$ is the partial solution in iteration $i$ after combining the $j$-th element from $w$. The changes introduced by the element $d$ are directly reflected in the partial solution after the microstep. Algorithm MICRO of Table 1 shows the structure of a microstep iteration. The iteration state $s$ and the working set $w$ are both incrementally updated by looking at one element $d \in w$ at a time. Similar to superstep iterations, microstep iterations are guaranteed to converge, if each individual update to the partial solution leads to a successor state in the CPO. Note that this is a stricter condition than for incremental iterations, where all updates together need to produce a successor state.

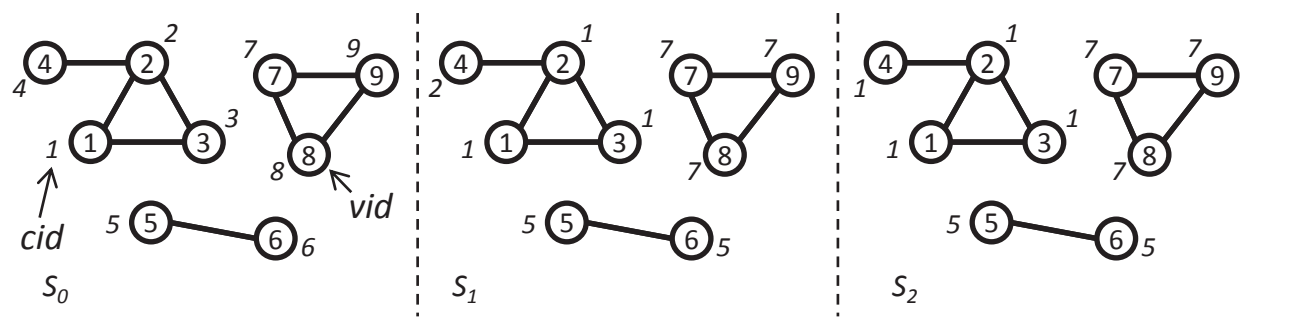

**Figure 1: Sample graph for the Connected Components.**

*Example: Connected Components.* Consider the pseudocode for the Connected Components algorithm shown in Algorithm MICRO-CC of Table 1. Inside each iteration, instead of performing two loops to update the state and the working set, these are simultaneously updated in one loop over the elements of the working set.

Note that in parallel setups, this last form of fixpoint iterations is amenable to asynchronous execution. The conformance of microsteps to the CPO can therefore enable fine-grained parallelism where individual element updates take effect in parallel, and no synchronization is required to coordinate the iterations/supersteps across parallel instances.

## 2.3 Performance Implications

The efficiency of bulk and incremental iterations may differ significantly. We illustrate this using the example of the Connected Components algorithm. In the bulk iteration algorithm, each vertex takes in each iteration the minimum component ID ($cid$) of itself and all its neighbors. Consequently, the number of accesses to the vertex state and the neighbor set is constant across all iterations. For the incremental (and microstep) variant of the algorithm, the cost of an iteration depends on the size of its working set.

Consider the sample graph from Figure 1. The numbers inside the vertices denote the vertex ID ($vid$), and the number next to a vertex denotes its current component ID ($cid$). The figure shows how the assigned $cid$ values change for each vertex over the three iterations. We can see that all except the vertex with $vid = 4$ reach their final $cid$ in one step. For most vertices, all their neighbors reach their final $cid$ in the first step as well. Those vertices need not be re-inspected. Their state cannot possibly change, since none of their neighbors' state did.

The incremental variant of the Connected Components algorithm reflects that by accessing only vertices, for which the working set contains new candidate $cid$s. That way, the algorithm focusses on "hot" portions of the graph, which still undergo changes, while "cold" portions are not accessed. The magnitude of this effect for a larger graph is illustrated in Figure 2. The graph is a small subset of the *Friend-of-a-Friend* network derived from a Billion-Triple-Challenge Web-Crawl and contains 1.2 million vertices and 7 million edges. The figure shows how the number of accesses and modifications to elements of the vertex states $s$ (left y axis), as well as the number of records added to the working set (right y axis) vary across iterations.

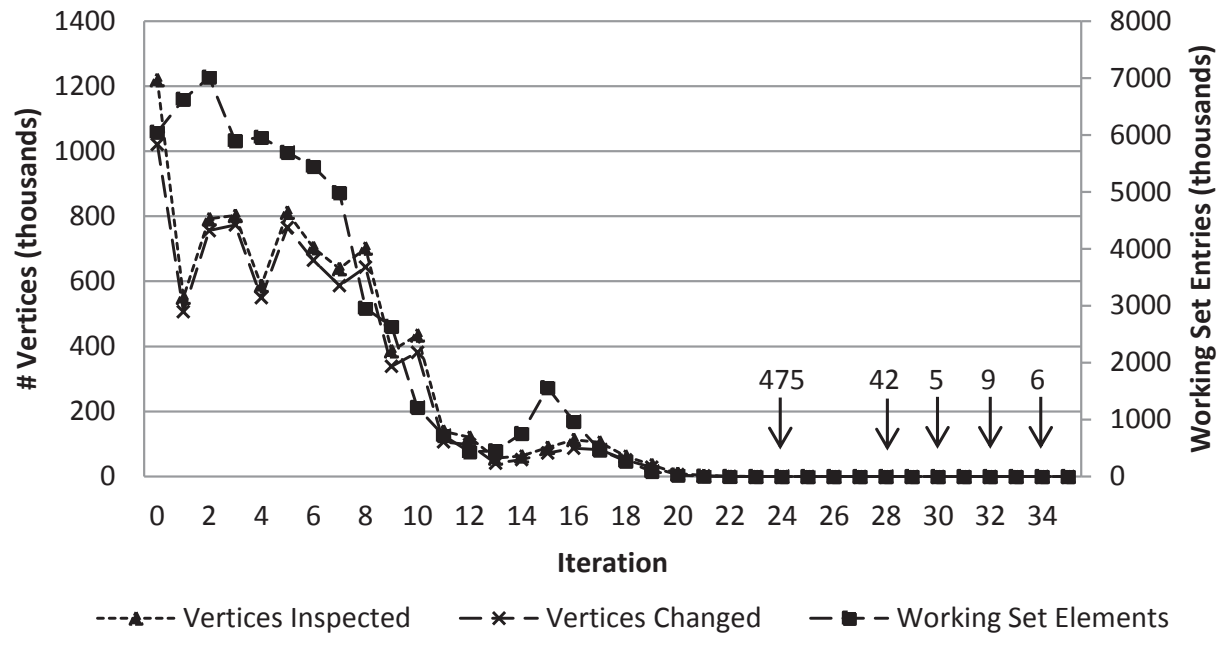

**Figure 2: The effective work the Connected Components algorithm performs on the FOAF subgraph.**

We see that the work performed in later iterations is significantly lower than the work in the first iterations, and that the actual progress (number of changed vertices) closely follows the size of the working set.

# 3. DATAFLOW SYSTEMS

We have implemented full and incremental iterations in the Stratosphere dataflow system [7]. However, the results presented in this paper are general, and can be applied in a variety of parallel dataflow systems, including parallel RDBMSs [17, 19], Hyracks [11], and SCOPE/Dryad [15]. This section briefly reviews the needed concepts from dataflow processing.

A dataflow is a directed acyclic graph (DAG) that consists of operators, sources, and sinks. The data sources and sinks, as well as the intermediate data sets that flow through operators, are bags of records. The operators are the inner nodes in the DAG, and can be thought of as functions $f : \{I_1, \ldots, I_n\} \rightarrow O$, where $I_i$ and $O$ are bags of records. A dataflow construct of several operators is therefore a function, whose fixpoint we can find by "closing the loop" in the dataflow DAG.

We do not make any assumptions about the semantics of the operators; indeed, operators can contain arbitrary user-defined code as in Stratosphere. It is interesting, however, to distinguish between certain classes of operators. First, we distinguish between operators that produce output by consuming one record, called *record-at-a-time* operators, and operators that need to consume multiple records before producing output, called *group-at-a-time* operators. In the popular MapReduce paradigm [16], an operator that implements the Map second-order function is a tuple-at-a-time operator, whereas Reduce operators are group-at-a-time; the latter need to consume all tuples in the incoming data set with a certain key value before producing output. A further distinction is between unary operators and operators that receive multiple inputs.

All dataflow systems include common data manipulation patterns, such as filtering based on predicate evaluation, joining of two datasets, and grouping a dataset according to the values of an attribute. In Stratosphere, user-defined code is encapsulated in so-called Parallelization Contracts (PACTs) [7]. PACTs are second-order functions that accept as arguments a user-defined first-order function, typically written in Java, and one or more input data sets. The type of PACT an operator implements informs the system about the possible distribution strategies of the operator's input. A Map PACT dictates that every record of the input $I$ can be processed independently. A Reduce PACT dictates that all records of $I$ with the same value of a key attribute are processed as a group. A Cross PACT produces an independent group from every pair of records of its input data sets, resembling a Cartesian product. A Match PACT groups pairs of records from two inputs only if the records have equal values on a key attribute, resembling an equi-join. Finally, a CoGroup PACT is a binary generalization of the Reduce contract, creating a group from all records of two inputs for every value of a key attribute.

Many dataflow systems, including Stratosphere, use an optimizer that decides on the execution strategy for the dataflow operators. For example, consider an equi-join, implemented via a Match contract. The Stratosphere optimizer here explores possible parallelization strategies, including broadcasting one input, partitioning one input, or re-partitioning both. Possible implementations include various flavors of hash-based or sort-merge-based execution [32]. The optimizer uses a cost model and interesting properties to generate an effcient plan for a given dataflow. Finally, we distinguish between pipelined operators and operators that materialize their input, and refer to the latter as materialization points or dams.

# 4. BULK ITERATIONS

This section describes the integration of bulk iterations into parallel dataflows. Bulk iterations recompute the entire partial solution in each iteration.

## 4.1 Dataflow Embedding

We represent this general form of iterations as a special construct that is embedded in the dataflow like a regular operator. An iteration is a complex operator defined as a tuple $(G, I, O, T)$. $G$ is a data flow that represents the step function $f : S \rightarrow S$, $S$ being the data type of the partial solution. The partial solution corresponds to the pair $(I, O)$, where $I$ is an edge that provides the latest partial solution as input to $G$. $O$ is the output of $G$, representing the next partitial solution.[2] In each but the first iteration, the previous iteration's $O$ becomes the next iteration's $I$. The iteration is embedded into a dataflow by connecting the operator providing the initial version of the partial solution $I$, and the operator cosuming the result of the last iteration to $O$. $T$, finally, denotes the *termination criterion* for the iteration. $T$ is an operator integrated into $G$ and is similar to a data sink in that it has only a single input and no output. It contains a Boolean function that consumes the input and returns a flag indicating whether to trigger a successive iteration. Instead of a termination criterion, the number of iterations $n$ may be statically defined. The iteration then is represented as a tuple $(G, I, O, n)$.

*Example: PageRank.* The PageRank algorithm [31] finds the fixpoint $p = A \times p$, where $p$ is the rank vector and $A$ is the left stochastic matrix describing the probabilities $p_{i,j}$ of going from page $j$ to page $i$. We represent the rank vector as a set of tuples $(pid, r)$, where $pid$ is the row index (and a unique identifier for the page), and $r$ is the rank. The sparse stochastic matrix $A$ is represented as a set of tuples $(tid, pid, p)$, where $tid$ is the row index (the unique identifier of the target page), $pid$ is the column index (the unique identifier of the source page), and $p$ is the transition probability. This representation does not include the tuples with $p = 0.0$. In each iteration the algorithm first joins the vector and the matrix tuple sets on $pid$, returning for each match $(tid, k = r * p)$. Second, all values are grouped by $tid$, which becomes the $pid$ for the result vector, and all $k$ are summed up to form the new rank $r$. Figure 3 shows the algorithm as an iterative dataflow. The big dashed box represents the iteration construct. Here, the dataflow $G$ comprises the rightmost Match operator (which joins the vector and the matrix), the Reduce operator (which groups and recomputes the rank), the data source $A$, and their connecting edges. The termination criterion $T$ uses another Match operator between the new and old rank data sets. The operator emits a record if a page's rank changed more than a threshold $\epsilon$.

[2]The concept extends straightforwardly to multiple data sets and hence multiple pairs $(I, O)_i$. For ease of exposition, we present the unary case in our examples.

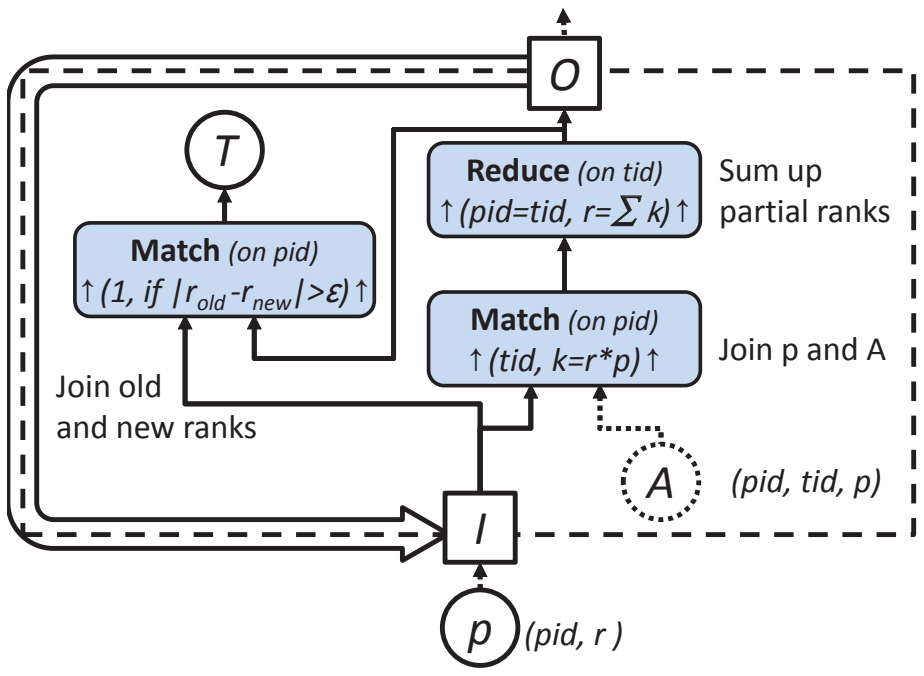

**Figure 3: PageRank as an iterative dataflow.**

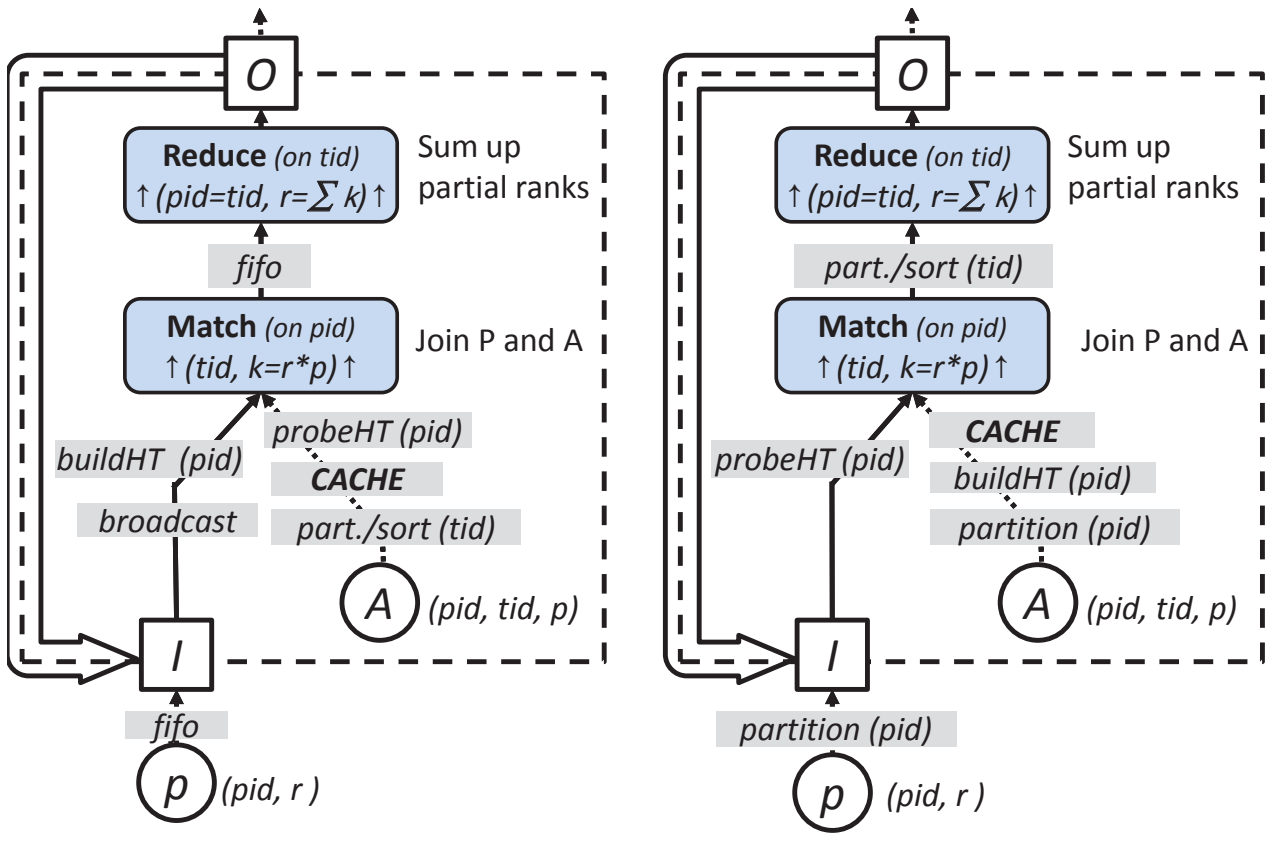

**Figure 4: Different execution plans for the PageRank iterative dataflow. The gray boxes show the optimizer's strategy choices.**

All nodes and edges in $G$ on the path from $I$ to $O$ and from $I$ to $T$ process different data during each iteration. We call that path the *dynamic data path*. In contrast, all nodes and edges that belong to the *constant data path* process data that is constant across all iterations. In Figure 3, the dynamic data path consists of the solid edges and the Match and Reduce operators. The constant data path includes data source $A$ and the edge connecting $A$ to the Match operatator. Note that data sources defined in $G$ are always on the constant data path.

## 4.2 Execution

Executing an iterative dataflow is possible by *"unrolling the loop"* or via *feedback channels*. When executing by unrolling, we roll out the iterations lazily: a new instance of $G$ is created whenever $O$ receives the first record and $T$ has signaled that another iteration is needed. The new instance of $G$ is concatenated to the current data flow by connecting the existing unconnected $O$ to the new $I$. During an iteration, $O$ may receive records before $T$ has consumed all records, depending on which operators in $G$ materialize their intermediate result. The PageRank example in Figure 3 describes such a case, as the Reduce operator emits records simultaneously to $O$ and the Match that is input to $T$. Here, an extra dam must be added to $O$, preventing the next iteration from receiving records before the decision whether to have that iteration was made. In some cases, the operator succeeding $I$ materializes its corresponding input (e.g. in a sort buffer or hash table). In that case, this specific materialization point serves as the dam and no extra dam is needed.

The feedback-channel based execution reuses $G$ in each iteration. Each operator in $G$ is reset after producing its last record. Similar as in the "unrolling" execution, the feedback edge materializes an iteration's result if $O$ receives records before $T$ decides upon a new iteration. Furthermore, if the dynamic data path contains less than two materializing operators, the feedback channel must also dam the dataflow to prevent the operators from participating in two iterations simultaneously. For PageRank in Figure 3, the feedback channel needs an additional dam if either the Reduce is pipelined or the right hand side Match pipelines its input $I$.

For all massively parallel systems, resilience to machine failures is necessary to ensure progress and completion of complex dataflows spanning many machines. Iterative dataflows may log intermediate results for recovery just as non-iterative dataflows, following their normal materialization strategies. In Stratosphere, for example, the Nephele execution engine judiciously picks operators whose output is materialized for recovery, trading the cost of creating the log against the potential cost of having to recompute the data. When executing with feedback channels, a new version of the log needs to be created for every logged iteration.

## 4.3 Optimization

Many dataflow systems optimize dataflow programs before executing them [7, 9, 17]. The optimization process typically creates an *execution plan* that describes the actual execution strategy for the degrees-of-freedom in the program. Such degrees-of-freedom comprise operator order, shipping strategies (partitioning, broadcasting) and local strategies (e. g., hashing vs. sorting operator implementations, as well as inner and outer role).

The optimizer may choose a plan for the iteration's data flow $G$ following the techniques for non-iterative dataflows [20, 33]. Note that in the general case, a different plan may be optimal for every iteration, if the size of the partial solution varies. The number of iterations is often not known a priori, or it is hard to get a good estimate for the size of the partial solution in later iteration steps. It is therefore hard for cost-based optimizers to estimate the cost of the entire iterative program. In our implementation in Stratosphere, we resort to a simple heuristic and let the optimizer pick the plan that has the least cost for the first iteration. For programs where the result size is rather constant across all iterations, that plan should be close to the best plan. To avoid re-executing the constant path's operators during every iteration, we include a heuristic that caches the intermediate result at the operator where the constant path meets the dynamic path. The caches are in-memory and gradually spilled in the presence of memory pressure. The cache stores the records not necessarily as an unordered collection, but possibly as a hash table, or $B^+$-Tree, depending on the execution strategy of the operator at the dataflow position where the constant and dynamic data paths meet. Finally, when comparing different plans, we weigh the cost of the dynamic data path by a factor proportional to expected number of iterations, since it is repeatedly executed. Plans that place costly operations on the constant data path are consequently cheaper than plans that place those operations on the dynamic data path.

Figure 4 shows two different execution plans for the PageRank algorithm, as chosen by Stratosphere's optimizer depending on the sizes of $p$ and $A$. The gray shaded boxes describe the optimizer's choices for execution strategies. It is noteworthy that the two plans resemble two prominent implementations of PageRank in Hadoop

MapReduce [4]. The left variant, as implemented by Mahout [5] and optimized for smaller models, replicates the rank vector and creates a hash table for it. This variant avoids to repeatedly ship the transistion matrix by caching it in partitioned and sorted form, such that grouping happens directly on the join result without additional partitioning/sorting. The right hand side variant, which is close to the PageRank implementation in Pegasus [26], joins a partitioned vector with the transition matrix and re-partitions for the aggregation. The transition matrix is here cached as the join's hash table. While in MapReduce, different implementations exist to efficiently handle different problem sizes, a dataflow system with an optimizer, such as Stratosphere, can derive the efficient execution strategy automatically, allowing one implementation to fit both cases.

Many optimizers in dataflow systems follow the Volcano approach [20], generating Interesting Properties (IPs) while they enumerate execution plan alternatives. Interesting Properties are used during plan pruning to recognize plans in which early operators can be more expensive, but produce data in a format that helps later operators to be executed more efficiently. For finding the optimal plan for $G$ across multiple iterations, the IPs propagated down from $O$ depend through the feedback on the IPs created for $I$, which themselves depend on those from $O$. In general, for an edge $e$ that is input to operator $P$, its interesting properties are $\text{IP}_e = \text{IP}_{P,e} \cup \text{AP}_e$, where $\text{IP}_{P,e}$ are the IPs that $P$ creates for that edge and $\text{AP}_e \subseteq \bigcup_{f \succ e} \text{IP}_f$ are the inherited properties, where $f \succ e$, if edge $f$ is a successor to edge $e$ in the DAG $G$. Note that $\text{IP}_{P,e}$ only depends on the possible execution strategies for $P$. A Match creates for example "partitioning" or "replication" as IPs for both edges. Which IPs are inherited depends on which properties could be preserved through the possible execution strategies of $P$ and the user code executed inside the operators[3]. The formula can be expanded to $\text{IP}_e \subseteq \bigcup_{f \succ e, P \in G} \text{IP}_{P,f}$. In the iterative setting, all edges on the dynamic data path are successors to all other edges, so an edge's interesting properties depend on all operators on the dynamic data path. To gather all relevant IPs for each edge, the optimization performs two top down traversals over $G$, feeding the IPs from the first traversal back from $I$ to $O$ for the second traversal.

In contrast to the methods originally described in [20], we interpret the interesting properties of an edge additionally as a hint to create a plan candidate that establishes those properties at that edge. The left hand plan in Figure 4 is the result of that procedure: the edge connecting $A$ and the Match operator has an interesting property for partitioning and sorting, generated by the Reduce operator. The plan candidate that applies partitioning and sorting at that point is actually very cheap in the end, because the expensive partitioning and sorting occur early on the constant data path.

# 5. INCREMENTAL ITERATIONS

In this section we discuss how to integrate incremental iterations, as described in Section 2, into dataflows.

## 5.1 Dataflow Embedding

An incremental iteration can be expressed using the bulk iterations introduced in Section 4, with two data sets ($S$ and $W$) for the partial solution and a step functions combining $u$ and $\delta$. The step function reads both data sets and computes a new version of $S$ and $W$. However, recall that the primary motivation for incremental iterations is to avoid creating a completely new version of the partial solution, but to apply point updates instead. The updated partial solution should be implicitly carried to the next iteration.

In imperative programming, updating the partial solution is achievable by modifying the statement $S = u(S, W)$ to $u(\&S, W)$, i. e., passing a reference to the state of the partial solution and modifying that shared state. Dataflow programs (like functional programs) require that the operators/functions are side effect free[4]. We work around this obstacle by modifying the update function from $S = u(S, W)$ to $D = u(S, W)$. The *delta set* $D$ contains all records that will be added to the partial solution and the new versions of the records that should be replaced in the partial solution. The solution set $S$ is treated as a set of records $s$, where each $s \in S$ is uniquely identified by a key $k(s)$. The delta set is combined with the solution set as $S = S \dot\cup D$. The $\dot\cup$ operator denotes a set union that, when finding two records from $S$ and $D$ with the same key, chooses the record from $D$: $S \dot\cup D = D \cup \{s \in S : \neg\exists d \in D | k(d) = k(s)\}$

We hence express an update of a record in the partial solution through the replacement of that record. The incremental iterations algorithm becomes

```
function INCR(δ,u,S,W)
  while W ≠ ∅ do
    D ← u(S,W)
    W ← δ(D,S,W)
    S = S ∪̇ D
```

Because the update function $u$ and the working set function $\delta$ frequently share operations, we combine them both to a single function $\Delta$, for ease of programmability: $(D_{i+1}, W_{i+1}) = \Delta(S_i, W_i)$

*Example: Connected Components.* The example follows the algorithm INCR-CC in Table 1. The solution set $S$ is a set of pairs $(vid, cid)$, which represents the mapping from vertex $(vid)$ to component ID $(cid)$. The $vid$ acts as the key that uniquely identifies a record in $S$. The working set $W$ contains pairs $(vid, cid)$. Each pair in $W$ is a candidate component ID $cid$ for vertex $vid$. Figure 5 shows the algorithm as an incrementally iterative dataflow. The dotted box represents the iteration operator, containing the dataflow for $\Delta$ that computes $W_{i+1}$ and $D_{i+1}$ from $W_i$ and $S_i$. We compute $D$ through an InnerCoGroup[5] operator that uses $vid$ as key. The InnerCoGroup calls its UDF for each $vid$ individually, providing the current $cid$ for the vertex from the input $S$, and all candidate $cid$s from $W$. Among the candidates, it selects the minimal $cid$ and, if that $cid$ is smaller than the operators current $cid$, it returns a pair $(vid, cid)$ with that new $cid$. When $D$ is combined with $S$, each record in $D$ replaces the record with the same $vid$ in $S$, thereby effectively updating the associated $cid$ for a $vid$.

The next working set $W_{i+1}$ is computed though a single Match operator that takes the delta set $D$ and joins it with the data source $N$ that represents the neighborhood mapping. $N$ contains the graph's edges as $(vid_1, vid_2)$ pairs[6]. The Match joins $D$ and $N$ via $vid = vid_1$ and returns for each match a pair $(vid_2, cid)$. Each returned pair represents the changed vertex's new $cid$ as a candidate $cid$ for its neighboring vertex $vid_2$.

[3] Reference [7] describes *OutputContracts* to determine how the user code behaves with respect to data properties.

[4] An intransparent side effect would void the possibility of automatic parallelization, which is one of the main reasons to use dataflow programming for large scale analytics.

[5] The InnerCoGroup is like a CoGroup, except that, much like an inner join, it drops groups where the key does not exist on both sides.

[6] We assume an undirected graph here, such that $N$ contains for every edge $(vid_1, vid_2)$ also the symmetric $(vid_2, vid_1)$ pair.

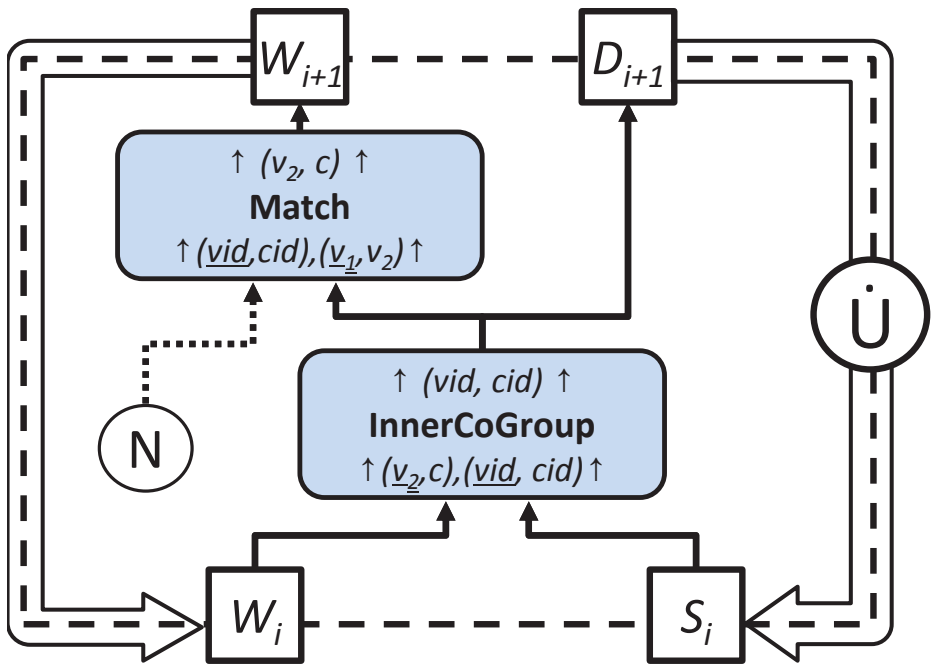

**Figure 5: The Connected Components algorithm as an Incremental Iteration.**

In fact, the iterative dataflow of Figure 5 can serve as a template for implementing a wide range of graph algorithms as incremental iterations. Every algorithm that can be expressed via a message-passing interface [3, 29] can also be expressed as an incremental iteration. $S(vid, state)$ represents the graph states, and $W(tid, vid, msg)$ represents the messages sent from vertex $vid$ to vertex $tid$. In each superstep, $\Delta$ combines the current state $S$ and messages $W$, it produces the changed states $D$, and creates the new set of messages using $D$ and possibly the graph topology table $N$.

By computing a delta set instead of the next partial solution, we can achieve that the iteration returns fewer records when fewer changes need to be made to the partial solution (cf. later iterations in Figure 2). The solution set $S$ is here a persistent state across the iterations, saving the cost of copying the unchanged records to the next iteration. Merging the small $D$ with $S$ is highly efficient if $S$ is indexed by the key that identifies the records.

To generalize the aforementioned example, we represent an *Incremental Iteration* as a complex dataflow operator, defined as a tuple $(\Delta, S_0, W_0)$. Let $S$ denote the *solution set*. $S$ is a set of records $s$ that are identified by a key $k(s)$. In iteration $i$, $S$ holds the $i$-th partial solution $S_i$. The initial version of $S$ is $S_0$, which is input to the iteration. After the incremental iteration has converged, $S$ holds the iteration's result. Let $W_i$ denote the working set for iteration $i$. The initial working set $W_0$ is input to the iteration.

The step function $\Delta$ computes in iteration $i$ the delta set $D_{i+1}$ with all new and changed records for $S_{i+1}$, as well as the working set $W_{i+1}$ for the next iteration: $(D_{i+1}, W_{i+1}) = \Delta(S_i, W_i)$. $\Delta$ is expressed as a dataflow in the same way as $G$ expresses the step function $f$ for bulk iterations (Section 4). Since $\Delta$ must return two data sets, it is necessarily a non-tree DAG. After $\Delta$ is evaluated for iteration $i$, $D_{i+1}$ is combined with $S_i$ using the modified union operator $\dot{\cup}$, producing the next partial solution $S_{i+1}$. That implies that any accesses to $S$ during the computation of $D_{i+1}$ and $W_{i+1}$ read the state of $S_i$. The iteration terminates once the computed working set is empty.

Because the delta set $D$ is the result of a dataflow operator, it is naturally an unordered bag of records, rather than a set. $D$ may hence contain multiple different records that are identified by the same key, and would replace the same record in the current partial solution. The exact result of $S \dot{\cup} D$ is then undefined. In practice, many update functions create records such that only one record with the same key can exist in $D$. That is, for example, the case in the Connected Components algorithm, because the InnerCoGroup operator joins the each record in the partial solution exactly once on the key that indexes it (the $vid$), and the UDF does not modify that field. Hence, each $vid$ appears at most once in the operators result. But since that is not necessarily the case in general, we allow the optional definition of a comparator for the data type of $S$. Whenever a record in $S$ is to be replaced by a record in $D$, the comparator establishes an order among the two records. The larger one will be reflected in $S$, and the smaller one is discarded. The usage of a comparator naturally reflects the strategy to establish an order among two states before and after a point update, as done in the definition of the CPO for $S$. Selecting the larger element represents the record leading to a successor state. Because the record from $D$ is dropped if it is the smaller one, $D$ relects only the records that contributed to the new partial solution.

## 5.2 Microstep Iterations

Section 2 discussed microstep iterations as a special case of incremental iterations. Recall that a microstep iteration is characterized by the fact that it takes a single element $d$ from the working set, and updates both the partial solution and the working set. Note that this implies that the partial solution already reflects the modification when the next $d$ is processed.

We represent microsteps iterations through the same abstraction as incremental iterations. In our implementation, an incremental iteration may be executed in microsteps rather than supersteps, if it meets the following constraints: first, the step function $\Delta$ must consist solely of record-at-a-time operations (e. g., Map, Match, Cross, ...), such that each record is processed individually[7]. For the same reason, binary operators in $\Delta$ may have at most one input on the dynamic data path, otherwise their semantics are ill defined. Consequently, the dynamic data path may not have branches, i. e. each operator may have only one immediate successor, with the exception of the output that connects to $D$. Note that this implies that $W_{i+1}$ may depend on $W_i$ only through $d$, which is consistent with the definition of microstep iterations in Table 1, line 5.

Finally, for microstep iterations, we need to assure that each time $\Delta$ is invoked, the partial solution reflects a consistent state, reflecting all updates made by prior invocations of $\Delta$. In a parallel system with multiple instances of $\Delta$ active at the same time, we encounter the classic transactional consistency problem, where guaranteeing a consistent up-to-date partial solution requires in the general case fine grained distributed locking of the elements in $S$ (or their keys). We can avoid the locking, when an update of a record in $S$ affects only the parallel instance that created the updated record. This is true, when the data flow between $S$ and $D$ does not cross partition boundaries, which is easily inferable from $\Delta$. The sufficient conditions are that the key field containing $k(s)$ is constant across the path between $S$ and $D$, and that all operations on that path are either key-less (e. g. Map or Cross) or use $k(s)$ as the key (for example in the case of the Match).

The Connected Components example above becomes amenable to microstep execution, if we replace InnerCoGroup operator by the record-at-a-time Match operator. Since its UDF keeps the key field constant, all records in $D$ will replace records in $S$ in the local partition.

## 5.3 Runtime & Optimization

The principal execution scheme for incremental iterations follows the bulk iteration scheme using feedback channels (Section 4.2). The techniques described for the partial solution in bulk iterations are used for the working set in the context of incremental iterations.

[7]For group- or set-at-a-time operations, supersteps are required to define the scope of the sets. Cf. systems for stream analytics, where windows or other reference frames are required to define set operations.

The dynamic and constant data path distinction applies directly to the $\Delta$ data flow in incremental iterations. The optimization, including caching static data and the modified handling of interesting properties happens the same way.

We extend those techniques to efficiently handle the persistent partial solution. To facilitate efficient access and updates to the partial solution, we store $S$ partitioned by its key across all nodes. Each node stores its partition of $S$ in a primary index structure, organized by the key. It is a typical pattern that the records from $S$ are joined (or otherwise associated) by an operator $o$ using the identifying key as the join key. In this case, we merge the $S$ index into $o$, creating a stateful operator that uses the $S$ index for its operation. The exact type of index is in this case determined by the execution strategy of $o$. For example, if $o$ is a join operator and the optimizer chooses to execute that join with a hash strategy, then $S$ will be stored in an updateable hash table. In contrast, if the optimizer picks a sort-based join strategy, $S$ is stored in a sorted index ($B^+$-Tree). That way, both accesses to $S$ by the $o$ and record additions to $S$ happen through the same index.

In the general case, we cache the records in the delta set $D$ until the end of the superstep and afterwards merge them with $S$, to guarantee a consistent view of $S$ during the superstep. Under certain conditions, the records can be directly merged with $S$, because we can guarantee that they will not be accessed in the same superstep again. Those conditions are equivalent to the conditions that guarantee updates to the partial soltution to be local.

Figure 6 illustrates the resulting execution strategy for the Connected Components algorithm, as derived by the optimizer. The working set $W$ is cached in queues, partitioned by $vid$ for the subsequent Match with $S$. $S$ is stored in a partitioned hash table using $vid$ as the key. When the Match between $W$ and $S$ returns a record, the plan writes the record back to the hash table. Simultaneously, it passes the record to the second Match function, which creates the new $(vid, cid)$ pairs for $W$. Note that the second Match uses the same key as the first Match, so it is not necessary to repartition the records by the join key. The contents of the data source $N$ is cached in a partitioned way as a hash table, such that $\Delta$ becomes pipelined and local. The plan partitions $\Delta$'s result (the new working set) and adds the records to the corresponding queues.

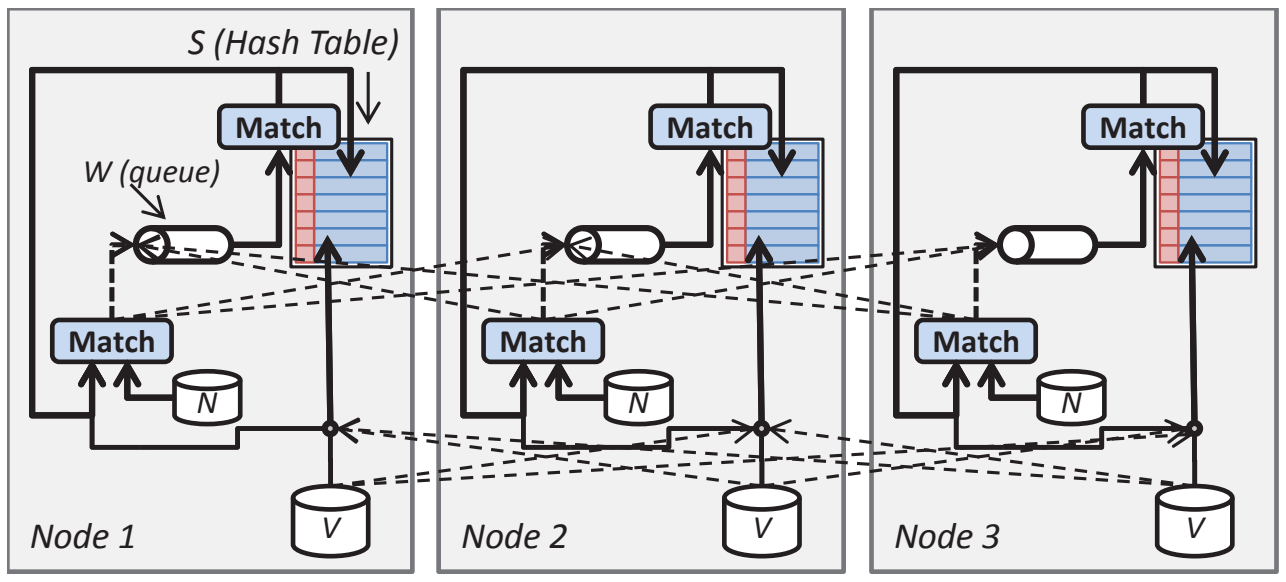

**Figure 6: The execution of the incrementally iterative Connected Components algorithm.**

If an iteration is amenable to execution in microsteps, the difference between superstep and microstep execution manifests only in the behavior of the queues that store $W$: for an asynchronous microstep iteration, they behave like regular nonblocking queues, passing records through in a FIFO fashion. In the presence of supersteps, the queues buffer all records that are added to them, but do not yet return them, as those added records are destined for the next iteration's working set. When all queues have returned all records for the current iteration, the superstep ends. The queues are then signaled to switch their buffers in a synchronized step, making the records added in the last iteration available to the current and buffering the records added in the current iteration for the next one. In our implementation in Stratosphere, we currently make use of the channel events offered by the Nephele dataflow engine to coordinate superstep transitions. Special channel events are sent by each node to signal the end of its superstep to all other nodes. Upon reception of an according number of events, each node switches to the next superstep.

In the synchronous case, the execution has converged if the working set $W_i$ is empty at the end of iteration $i$. A simple voting scheme can here decide upon termination or progression. For the asynchronous execution, however, no end-of-superstep point exists. The distributed systems community has explored a variety of algorithms for termination detection in processor networks. Many of these algorithms apply also to the asynchronous execution of the parallel dataflow. For example, [27] works by requesting and sending acknowledgements for the records along the data channels.

# 6. EVALUATION

To evaluate the practical benefit of incremental iterations, we compare three systems that support iterations in different ways: *Stratosphere* [7], *Spark* [36], and *Giraph* [3].

*Stratosphere* supports both bulk and incremental iterations, which were implemented as described in Sections 4 and 5. The implementation uses the feedback-channel based execution strategy.

*Spark* is a parallel dataflow system implemented in Scala and centered around the concept of Resilient Distributed Data Sets (RDSs). RDSs are partitioned intermediate results cached in memory. Spark queries may contain operations like Join, Group, or CoGroup and apply user defined functions on the operators results. The system accepts iterative programs, which create and consume RDSs in a loop. Because RDSs are cached in memory and the dataflow is created lazily when an RDS is needed, Spark's model is well suited for bulk iterative algorithms.

*Giraph* is an implementation of Google's Pregel [29] and hence a variant of Bulk Synchronous Parallel processing adopted for graph analysis. The model is explicitly designed to exploit sparse computational dependencies in graphs. A program consists in its core of a vertex update function. The function computes its update based on messages it receives from other vertices, and sends out messages to other vertices in turn. Because the function has only a local view of one vertex, the algorithms have to be expressed by means of localizing the updates. Pregel is thus a special case of incremental iterations - the vertices represent the partial solution state and the messages form the working set.

All of the above systems run in a Java Virtual Machine (JVM), making their runtimes easy to compare. We ran our experiments on a small cluster of four machines. Each machine was equipped each with 2 Intel Xeon E5530 CPUs (4 cores, 8 hardware contexts) and 48 GB RAM. The machines' disk arrays read 500 MB/sec, according to `hdparm`. We started the JVMs with 38 GB heap size, leaving 10 GB to operating system, distributed filesystem caches and other JVM memory pools, such as for native buffers for network I/O. The cluster has consequently 32 cores, 64 threads and an aggregate Java heap size of 152 GB.

We use four different graphs as data sets, which we obtained from the University of Milan's Web Data Set Repository [10]: the link graph from the English Wikipedia and the Webbase web crawl from 2001 are typical web graphs. The Hollywood graph, linking Actors that appeared in the same movie, and the Twitter follower graph are representatives of social network graphs. The latter are typically more densely connected. Table 2 shows the graphs' basic properties.

| DataSet | Vertices | Edges | Avg. Degree |
|---|---|---|---|
| Wikipedia-EN | 16,513,969 | 219,505,928 | 13.29 |
| Webbase | 115,657,290 | 1,736,677,821 | 15.02 |
| Hollywood | 1,985,306 | 228,985,632 | 115.34 |
| Twitter | 41,652,230 | 1,468,365,182 | 35.25 |

**Table 2: Data Set Properties**

## 6.1 Full Iterations

We first establish a base line among the three systems using the bulk iterative PageRank algorithm, as described in Section 4. For Giraph and Spark, we used the implementations that were included with the systems' code. Giraph's algorithm follows the example in Pregel [29], and Spark's implementation follows Pegasus [26]. For Stratosphere, we executed both strategies from Figure 4. The partitioning plan (right hand side of the figure) is equivalent to the Spark implementation. The broadcasting plan (left hand side) is cheaper by network cost, because it computes the new ranks locally.

We run PageRank for 20 iterations. Even though computational dependencies do exist in the graphs, the algorithm operates in batches, updating each vertex every time. Consequently, we expect all systems to have roughly equal runtime for PageRank because all iterations do the same work: they create records for each edge propagating the current rank of a vertex to its neighbors. These records are pre-aggregated (cf. Combiners in MapReduce and Pregel) and are then sent over the network to the node containing the neighbor vertex. An exception is, as mentioned, Stratosphere's broadcasting strategy.

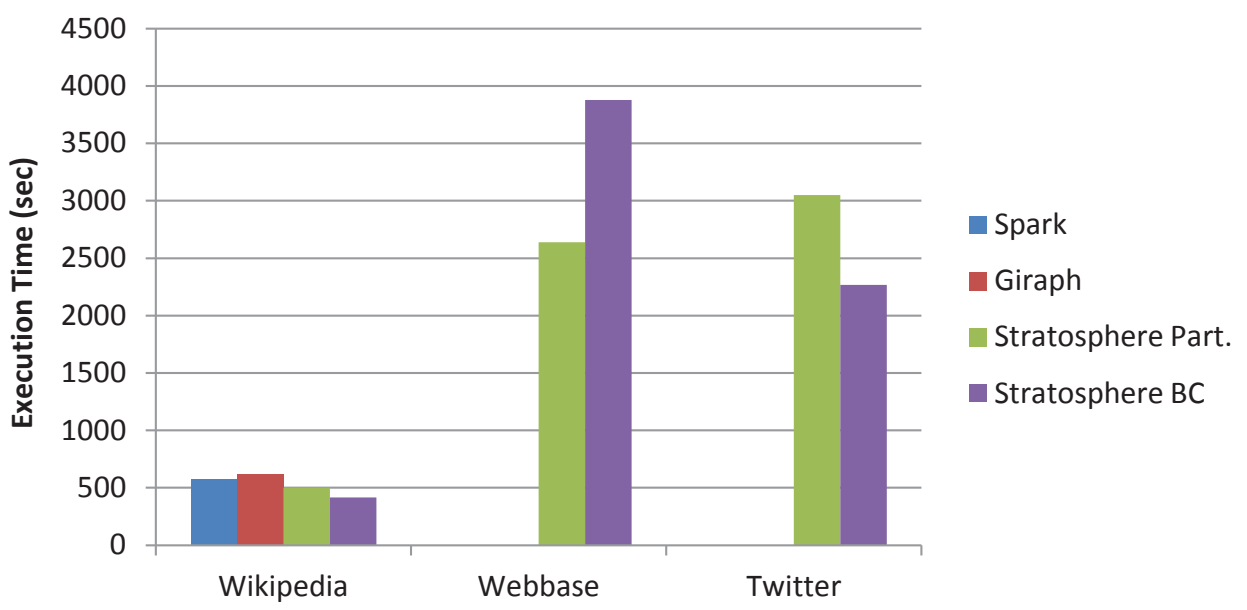

**Figure 7: Total execution times for the PageRank algorithm.**

Figure 7 shows the results for the PageRank algorithm on the three systems on different datasets. As expected, the runtime of the algorithm is similar in Spark, Giraph, and Stratosphere for the small Wikipedia dataset. We were unable to use Spark and Giraph with the large datasets, because the number of messages created exceeds the heap size on each node. The systems currently lack the feature to spill messages in the presence of memory pressure. For the large Webbase graph, we see that Stratosphere's broadcasting strategy degrades. Due to a limitation in the hash join algorithm, the hash table containing the broadcasted rank vector is currently built by a single thread on each machine, introducing a critical serial codepath.

It is interesting to examine the time each iteration takes in the different systems. Figure 8 breaks down the execution time for the PageRank algorithm on the Wikipedia dataset. For Stratosphere, we examine the partitioning strategy, because it performs the same work as the two other systems. The iteration times are rather constant in Stratosphere and Giraph. The first iteration is longer, because it includes for Stratosphere the execution of the constant data path, and for Giraph the partitioning and setup of the vertex states. The iteration times in Spark vary quite heavily due to garbage collection problems[8]. Spark uses new objects for all messages, creating a substantial garbage collection overhead. In contrast, Stratosphere's PACT runtime stores records in serialized form to reduce memory consumption and object allocation overhead. The runtime algorithms are optimized for operation on serialized data. The Giraph implementation is also hand tuned to avoid creating objects where possible, removing pressure from the garbage collector.

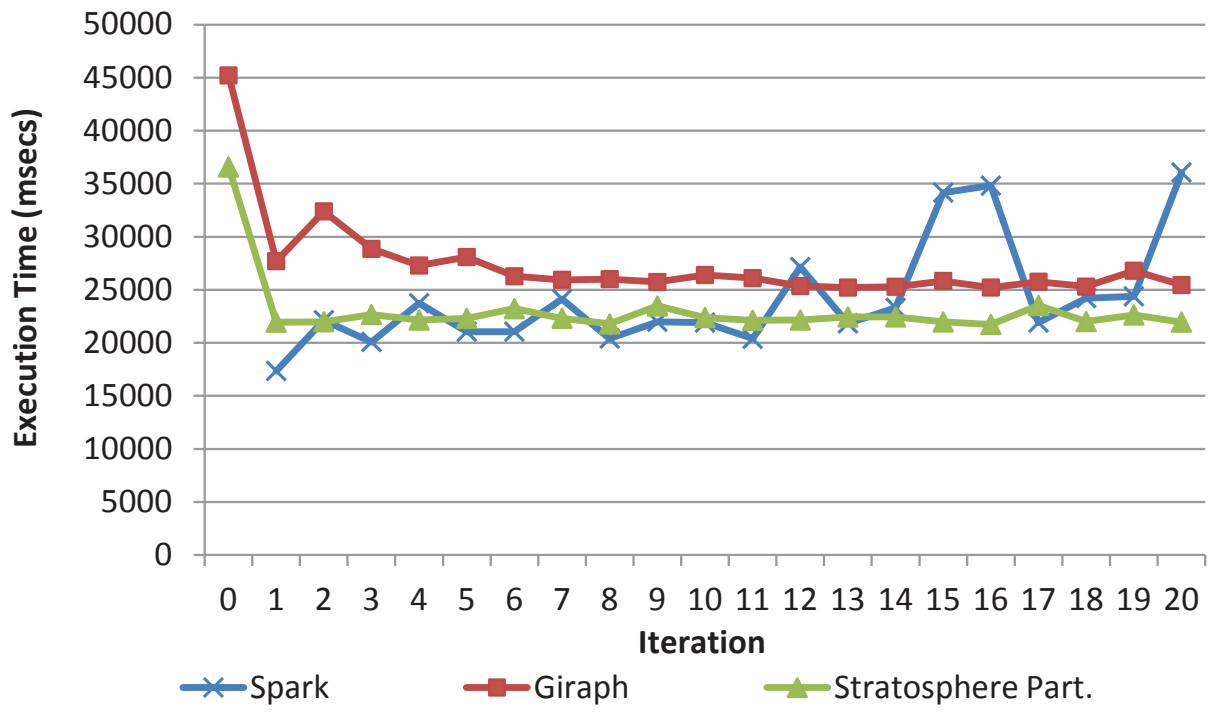

**Figure 8: Execution times of the individual iterations for the PageRank algorithm on the Wikipedia dataset.**

## 6.2 Incremental Iterations

While we expected the systems to perform similarly for bulk iterative algorithms, we expect a big performance difference for incrementally iterative algorithms. The magnitude of the difference should depend on the sparsity of the computational dependencies: the sparser the dependencies are, the more speedup we expect through incremental iterations, as compared to bulk iterations.

To verify our expectations, we ran the different versions of the Connected Components algorithm on the three systems. For directed graphs, we interpreted the links as undirected, thus finding weakly Connected Components. We ran the bulk iterative algorithm from Section 2 on Spark and Stratosphere and an the incrementally iterative version on Stratosphere and Giraph.

We have two Stratosphere implementations for the incrementally iterative algorithm, resembling the examples in Section 5: one uses a Match operator for the update function, the other one a CoGroup. Recall that the Match variant corresponds to a microstep iteration and takes each element from the working set individually, potentially updating the partial solution and producing new working set elements. The CoGroup variant represents a batch incremental iteration. It groups all working set elements relevant to one entry of the partial solution, such that each such entry is updated only once. The grouping of the working set does, however, incure some additional cost. We consequently expect the Microstep variant to be faster on the sparse graphs, where the number of redundant candidates in the working set is smaller. For denser graphs, we expect the batch incremental algorithm to eventually be the faster variant, as the cost of grouping the working set elements is amortized by the saved cost due to fewer accesses to the partial solution.

Figure 9 shows the execution times for the Connected Components algorithm. We were again unable to run the algorithm on Giraph and Spark for the Twitter and Webbase dataset, as the systems ran out of memory.

For the Wikipedia graph, the algorithms took 14 iterations to converge. The speedup of incremental iterations compared to bulk iterations in Stratosphere is roughly a factor of 2. Giraph also clearly outperforms the bulk iterations in Spark and Stratosphere.

[8]The iteration times for Spark were averaged from multiple runs of the algorithm. Inside a single run, the variance is even greater.

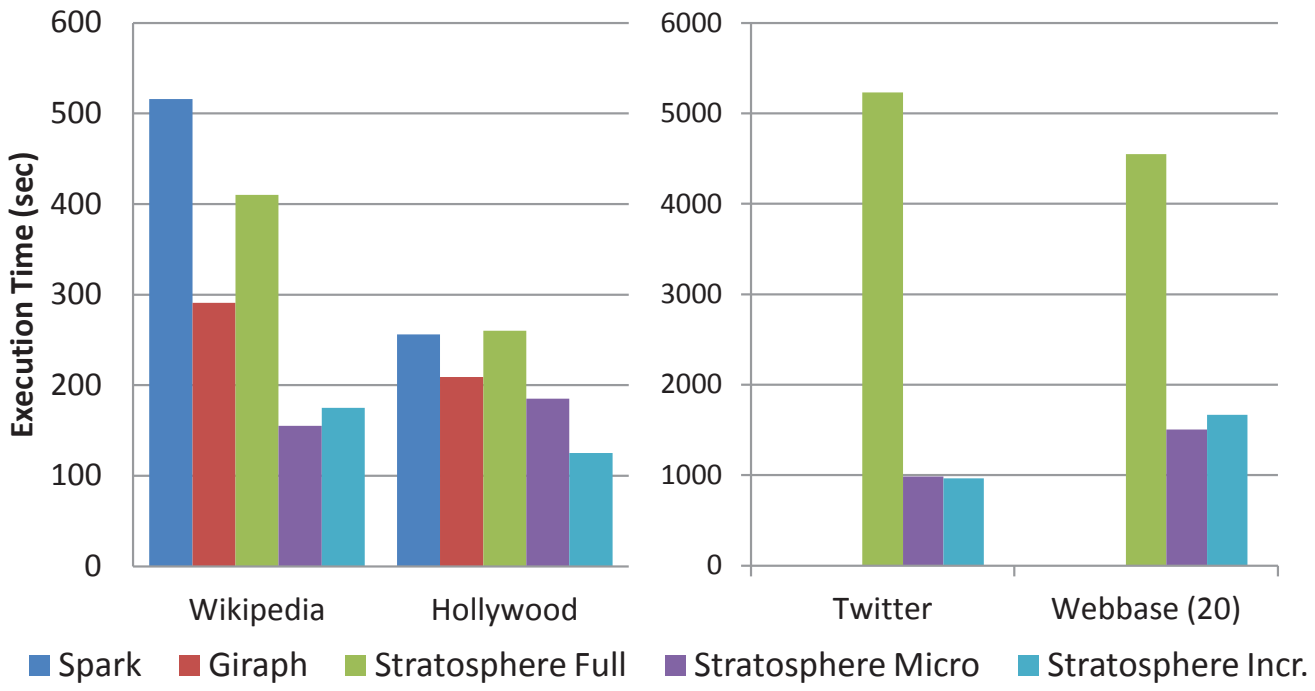

**Figure 9: Total execution times for the Connected Components algorithm.**

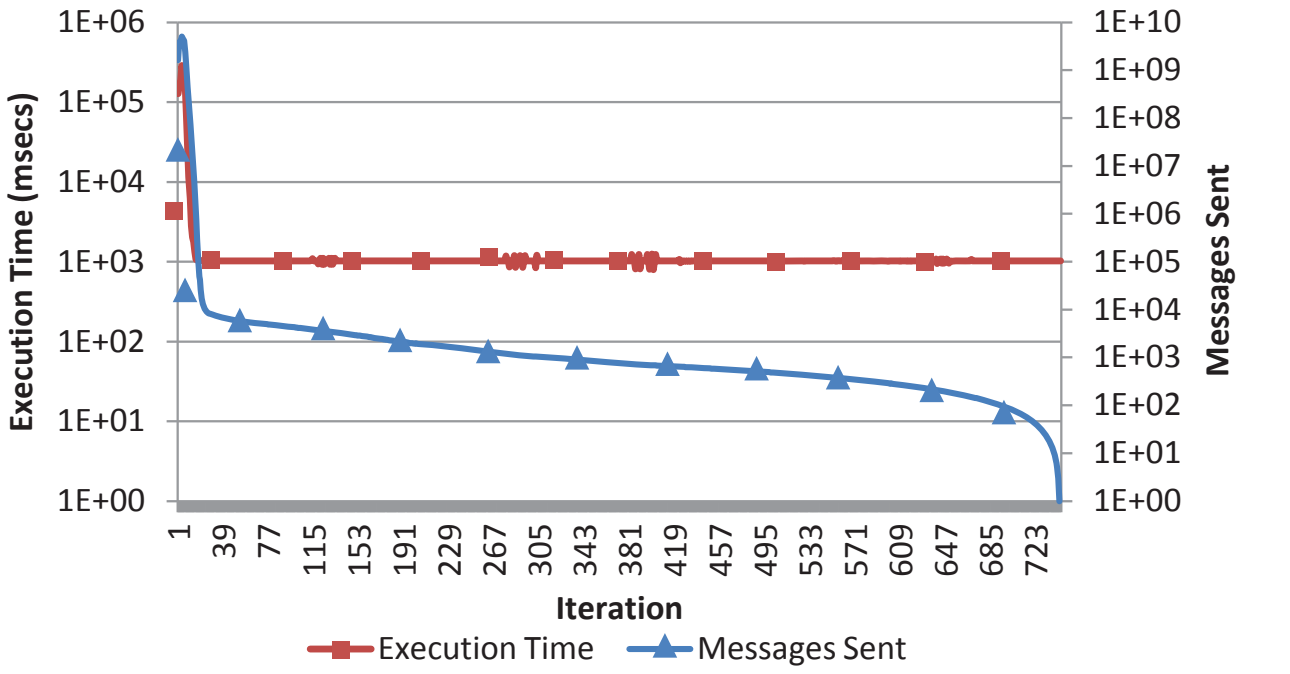

**Figure 10: Execution time and working set elements for Connected Components in the Webbase graph on Stratosphere.**

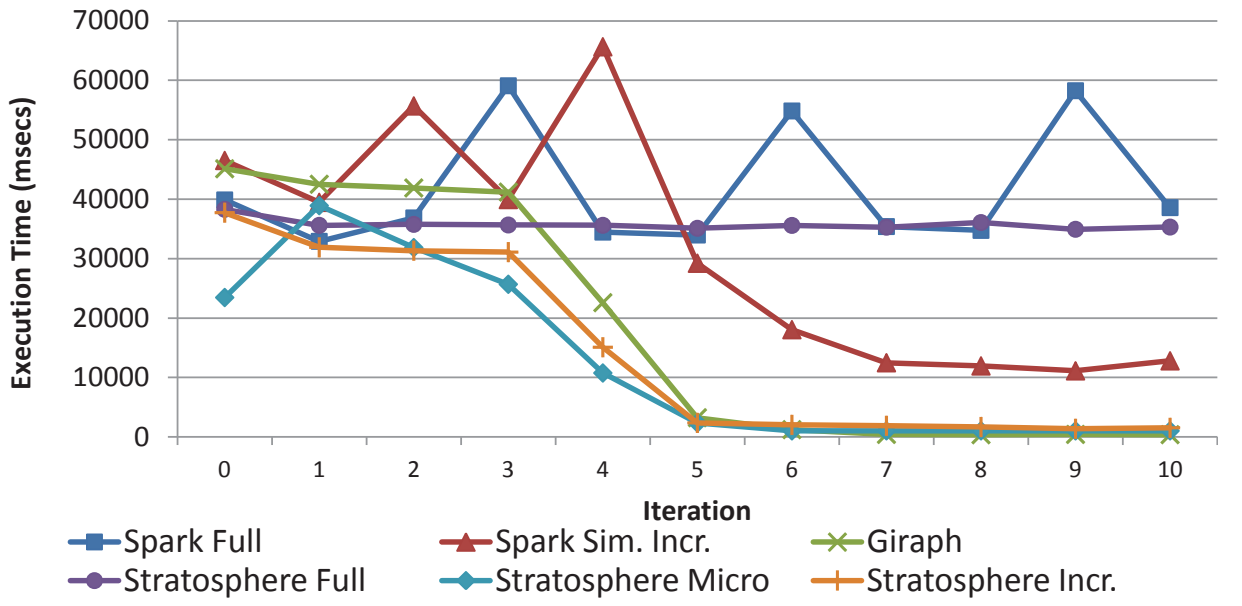

**Figure 11: Execution times of the individual iterations for the Connected Components algorithm.**

For the Hollywood data set, the gain though incremental iterations is less, because the vertex connections, and consequently the computational dependencies, are more dense. In contrast to the runtimes on the Wikipedia graph, the batch inremental algorithm is here roughly 30% faster than the microstep algorithm. As discussed before, this is due to the graph's density, which results for the microstep algorithm in many more accesses to the partial solution, while the batch incremental algorithm only groups a larger set.

For the Twitter dataset, where the algorithms takes 14 iterations to converge, we see a tremendous gain through incremental iterations. They are in Stratosphere roughly 5.3 times faster than the bulk iterations. The performance gain is high, because a large subset of the vertices finds its final component ID within the first four iterations. The remaining 10 iterations change less than 5% of the elements in the partial solution.

For the Webbase graph, the figure shows the algorithm runtimes for the first 20 iterations. Because the largest component in the graph has a huge diameter, the algorithms require 744 iterations to fully converge. Already in the first 20 iterations, in which the majority of the actual component ID changes happen, the incrementally iterative algorithms are 3 times as fast as their bulk iterative counterparts. For the entire 744 iterations, the incremental algorithms take 37 minutes, which is only a few minutes longer than for the first 20 iterations. In contrast, the extrapolated runtime for the bulk iterative algorithm to converge on the whole graph is roughly 47 hours. The comparison yields a speedup factor of 75. By the time the incremental algorithms converge on the entire graph, the bulk iterative algorithm has just finished its 10th iteration. Figure 10 shows the execution time and the number of working set elements per iteration for the Webbase graph. The execution time does not drop below 1 second, which is a lower bound currently imposed by synchronization of the steps.

In Figure 11 we examine the iteration times for the various algorithms processing the Wikipedia graph. To assess the benefit from sharing the partial solution state across iterations, we added an additional iteration variant that simulates an incremental iteration on top of Spark's non-incremental runtime. This simulated variant adds a Boolean flag to the each entry in the partial solution. The flag indicates whether the component ID decreased in the last iteration. If it did not, then no messages are sent to the neighbors, but only to the vertex itself, in order to carry the current component ID mapping to the next iteration. This simulated variant exploits computational dependencies, but needs to explicitly copy unchanged state.

We see that the bulk iterations in Stratosphere and Spark have a constant iteration time (modulo the variance introduced by garbage collection in Spark). In contrast, the incrementally iterative algorithms in Stratosphere and Giraph converge towards a very low iteration time after 4 iterations. They exploit the sparse computational dependencies and do not touch the already converged parts of the partial solution in later iterations. Giraph's iteration times are the lowest after 5 iterations, because it currently uses a more efficient way to synchronize iterations than Stratosphere. The simulated incremental algorithm on Spark (Spark Sim. Incr.) decreases in iteration time as well, but sustains at a much higher level, because of the cost for copying the unchanged records in the partial solution.

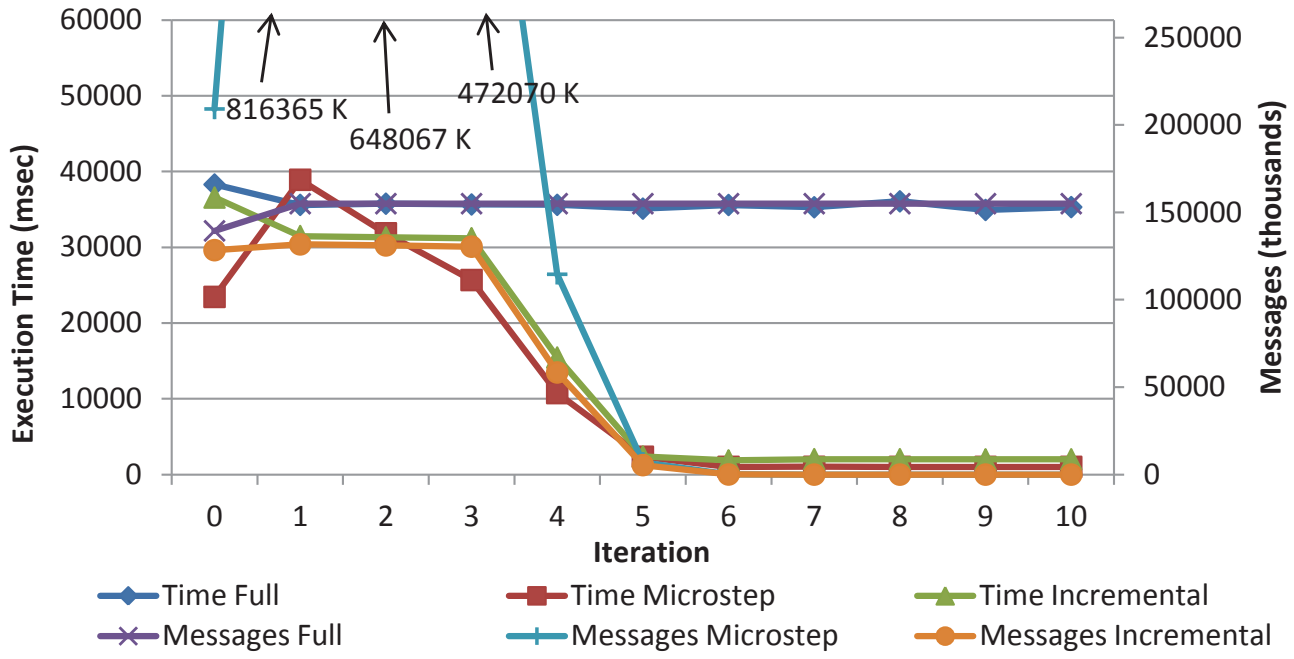

**Figure 12: Correlation between the execution time and the exchanged messages for the Wikipedia graph on Stratosphere.**

We finally look at how the number of created candidate component IDs (messages, respectively working set elements), influence the runtime of bulk iterative and incrementally iterative algorithms. Figure 12 shows for the Wikipedia data set the time per iteration and the number of messages (working set elements). By the overlapping curves, we see that for the bulk iterative algorithm and the batch incremental algorithm (CoGroup), the runtime is almost a linear function of number of candidate component IDs, in both cases with the same slope. For the microstep iterative algorithm (Match), a similar relationship exists but with a much lower slope. Because in the microstep variant, the update function is much cheaper than for the batch incremental variant, it can process in equal time much larger working sets containing many more redundant candidate component IDs.

We have seen that Stratosphere, as a general iterative dataflow system, can outperform specialized analysis systems for bulk iterative algorithms. At the same time, the incremental iterations speed up algorithms that exploit sparse computational dependencies. They touch in each iteration only the elements in the partial solution that potentially require modification, thereby greatly outperforming full iterations.

# 7. RELATED WORK

In this section, we first relate our work to recursive query evaluation, which is the most prominent alternative to iterative queries. Subsequently, we relate our work to other systems for parallel data analysis.

## 7.1 Relationship to Recursive Queries

In programming languages, the equally expressive alternative to iteration is recursion. There has been a wealth of research addressing recursion in the context of Datalog and the relational model [6]. The query processing line of that research focused largely on top-down versus bottom-up evaluation techniques and suitable optimizations, like predicate pushdown [8]. The incremental iteration techniques presented here bear similarity to the *semi-naïve* evaluation technique. More recently, Afrati et al. [2] and Bu et al. [12] discuss how to compile recursive Datalog queries to dataflows. Both evaluate the recursion bottom up. The approach of [2] uses stateful operators for the datalog rules, while [12] discusses a more comprehensive approach including optimization of the created dataflow. Our incremental iterations offer a similarly expressive abstraction. All examples from [2, 12] are expressible as incremental iterations. In fact, following the arguments from Afanasiev et al. [1] and Bu et al. [12], it becomes clear that one can express stratified (or XY-stratified) Datalog progams as incremental iterations with supersteps. For programs without negation and aggregation, the supersteps can be omitted. Since recursion in relational databases has much stricter conditions than in Datalog, the proposed incremental evaluation is applicable to relational DBMSs. Early work in that direction includes *delta iterations* [21]. We build on a similar abstraction, but additionally focus on aspects relevant to parallel execution.

Both Datalog queries (and recursive relational queries) may only add rule derivations (respectively tuples) to the partial solution. There is no synchronization barrier defined, so this model naturally lends itself to asynchronous execution. For algorithms that require supersteps, Bu et al. [12] express synchronization barriers by introducing a temporal variable that tracks the supersteps and using aggregation predicates to access the latest superstep's state. Because this leads easily to a complicated set of rules, they use this technique to specify domain specific programming models rather than to specify queries directly. A distinguishing aspect in our approach is the fact that incremental iterations can actually update the state of the partial solution, allowing the iterations to compute non-inflationary fixpoints (cf. [1]). This obliterates the need to filter the state for the latest version and naturally leads to a semi-naïve flavor of evaluation, resulting in simple queries.

In contrast to the work of [12], we do not distinguish between global- and local model training, but leave the decision to replicate or broadcast the model to the dataflow optimizer. Instead, we distinguish between iterations that recompute the partial solution or incrementally evolve it, because that has a direct impact on how the programmer has to specify the algorithm.

## 7.2 Other Iterative Analytical Systems

HaLoop [13] and Twister [18] are MapReduce [16] extensions for iterative queries. They provide looping constructs and support caching of loop-invariant data for the special case of a MapReduce dataflow. In contrast, our technique addresses more general dataflows. The optimization techniques presented in Section 4.3 subsume their special-case optimization. The Spark dataflow system [36] supports resilient distributed datasets (RDS). With loop constructs in its Scala front-end, it supports full iterations. Neither of the systems addresses incremental iterations or considers dataflow optimization.

Pregel [29] is a graph processing adoption of bulk synchronous parallel processing [34]. Programs directly model a graph, where vertices hold state and send messages to other vertices along the edges. By receiving messages, vertices update their state. Pregel is able to exploit sparse computational dependencies by having a mutable state and the choice whether to update it and propagate changes in an iteration. The incremental iterations proposed in this paper permit equally efficient programs, with respect to the number of times the state of a vertex is inspected, and the number of messages (in our case records) sent. It is straightforward to implement Pregel on top of Stratosphere's iterative abstraction: the partial solution holds the state of the vertices, the workset holds the messages. The step function $\Delta$ updates the vertex state from the messages in the working set and creates new update messages. Since $\Delta$ may be a complex parallel data flow in itself, it allows for writing even more complex algorithms easier. The adaptive version of PageRank [25] for example, can be expressed as an incremental iteration, while it is hard to express it on top of Pregel. The reason for that is that Pregel combines vertex activation with messaging, while incremental iterations give you the freedom to separate these aspects. In addition, incremental iterations have well defined conditions when they allow for (asynchronous) microstep execution, and offer a dataflow abstraction, thus integrating naturally with parallel dataflow systems.

GraphLab [28] is a framework for parallel machine learning, where programs model a graph expressing the computational dependencies. The abstraction is distributed shared memory: nodes exchange data by reading neighboring nodes state. GraphLab has configurable consistency levels and update schedulers, making it a powerful, but low level abstraction. Most GraphLab algorithms can be implemented as incremental iterations by using the working set to hold the IDs of scheduled vertices and accessing neighboring states though a join with the partial solution.

ScalOps [35] is a domain specific language for big data analytics and machine learning. It provides a simple language interface with integrated support for iterations. ScalOps programs are compiled to optimized Hyracks [11] dataflows. At the time of writing this paper, the system aspects of executing ScalOps on top of Hyracks have not been published.

Ciel [30] is a dataflow execution engine that supports very fine grained task dependencies and runtime task scheduling suitable for iterations and recursions. Ciel does not offer a direct dataflow abstraction to iterations or recursions; instead, the task dependencies are generated from queries in a specialized domain specific language. Ciel cannot share state across iterations, but the runtime task scheduling allows to somewhat mimic that, although at the cost of creating very many tasks.

## 8. CONCLUSIONS AND OUTLOOK

We have shown a general approach to iterations in dataflows. For algorithms with sparse computational dependencies, we argued that their exploitation is crucial, as it leads to a huge speedup. To exploit computational dependencies in dataflow systems, we suggested an abstraction for incremental iterations. Incremental iterations are general and suit many iterative algorithms. We presented optimization and execution strategies that allow a dataflow system to execute the incremental iterations efficiently, touching only the state that needs to be modified to arrive at the iteration's result.

Incremental iterations subsume several specialized analysis models, such as the Pregel model, both in expressibility and efficiency. In fact, all algorithms that are naturally expressible in Pregel's programming model, can be expressed as incremental iterations without a performance hit. Furthermore, incremental iterations allow for (asynchronous) microstep execution under well defined conditions.

In the case study with our prototypical implementation of incremental iterations in the Stratosphere system, we have shown that an implementation of incremental iterations allows a parallel dataflow system to compete with specialized systems for algorithms that are a sweet spot of those systems. At the same time, Stratosphere maintains a general dataflow abstraction, thus reducing the number of systems required for large scale analytics.

In future work we want to explore how far compiler techniques can automatically transform algorithms stated as fully iterative algorithms into incremental algorithms. Furthermore, we plan to investigate how fault tolerance techniques for full iterations can benefit from the fact that the iteration repeatedly executes the same data flow and may learn about its characteristics across iterations.

## 9. ACKNOWLEDGEMENTS

Stratosphere is funded by the DFG FOR 1306 grant. We would like to thank Joe Harjung for his comments on fixpoint functions and Sebastian Schelter for his help with Giraph.